\documentclass[aps,twocolumn,groupedaddress,nofootinbib]{revtex4}
\expandafter\let\csname equation*\endcsname\relax
\expandafter\let\csname endequation*\endcsname\relax
\usepackage{csquotes}
\usepackage{graphicx,color}
\usepackage{amsmath}
\usepackage{amssymb}
\usepackage{multirow}

\begin{document}

\title{Low energy electron reflection from tungsten surfaces}

\author{P. Tolias}
\affiliation{Division of Space and Plasma Physics, Association EUROfusion-VR, Royal Institute of Technology KTH, Stockholm, Sweden}


\begin{abstract}
\noindent The incidence of very low energy electrons on metal surfaces is mainly dictated by the phenomenon of quantum mechanical reflection at the metal interface. Low energy electron reflection is insignificant in higher energy regimes, where the more familiar secondary electron emission and electron backscattering processes are the dominant features of the electron-metal interaction. It is a highly controversial subject that has mostly emerged during the last years. In this brief note we examine the source of the controversy, present some basic theoretical considerations, recommend a dataset of reliable experimental results for the reflection of low energy electrons from tungsten surfaces and discuss the suppression of reflected electrons by external magnetic fields in the light of applications in fusion devices.
\end{abstract}

\maketitle

\noindent \emph{Motivation.} For relatively large incident electron energies $E_{\mathrm{inc}}^{\mathrm{e}}\gtrsim50\,$eV, electron emission from solids induced by electron impact is mainly a consequence of backscattering of primaries and emission of secondaries. Recently, reliable experimental results on electron backscattering and secondary electron emission from tokamak relevant materials have been reviewed and semi-empirical quasi-universal expressions have been proposed to describe the respective yields\,\cite{mypaper1,mypaper2}. We emphasize though that these expressions are valid for $E_{\mathrm{inc}}^{\mathrm{e}}\gtrsim50\,$eV and that extrapolations below this energy range should not be carried out. However, in lack of an alternative option, such extrapolations are carried out in many fusion works. For this reason, in this work we shall review theoretical and experimental results on electron induced electron emission in the regime of very low incident energies. Before proceeding, it is essential to define the inner metal potential that is the energy gained by a free electron when it penetrates a metal surface and is equal to the energy difference between the vacuum level and the bottom of the valence band, $V_0=B_{\mathrm{w}}+W_{\mathrm{f}}$, where $W_{\mathrm{f}}$ is the work function and $B_{\mathrm{w}}$ is the valence band width.

For $E_{\mathrm{inc}}^{\mathrm{e}}\lesssim50\,$eV, electron backscattering and secondary electron emission from metals start becoming negligible. Even more important, at such a low energy regime their semi-empirical description breaks down being largely based on the continuous slowing down approximation, since the incident electrons can only penetrate few atomic layers of the solid before thermalizing. Moreover, when the incident energy becomes comparable with the inner metal potential that acquires values within the range $\sim10-20\,$eV for most metals, reflection of the incoming electron at the potential barrier becomes significant. This phenomenon is typically coined as low energy electron reflection and it is quantified by the reflection coefficient $\xi_{\mathrm{e}}$, the ratio of reflected to incident electrons.

There has been a small number of experimental works claiming that electron reflection from technical metal surfaces at low incident energies can be significantly enhanced, even reach $100\%$ at the limit of zero incident energy\,\cite{yieldone}. This argument has been iterated within the fusion community in order to explain a number of experimental observations\,\cite{fusrefl1,fusrefl2,fusrefl3,fusrefl4}. However, it not only contradicts concrete experimental evidence for clean metal surfaces but also lacks a sound theoretical basis. For these reasons it has been recently criticized on both ends\,\cite{Migraine,critici1,critici2}. It is safe to state that the reflection coefficient cannot reach values close to unity, unless plasma fundamentally changes the nature of electron-solid interactions or the nature of the surface in a manner not documented at the moment.

\noindent \emph{Survey of theoretical results.} We begin our investigation of low energy electron reflection with presenting some theoretical considerations in order to demonstrate that the electron reflection coefficient does not reach unity but also in order to put emphasis to the intricate nature of the problem. We assume that the metal surface is planar and the electron incident velocity is normal to the surface. It is desirable to approximate the passage of an electron to the metal interior by a single-body description, where the one-electron Schr\"odinger equation is solved for an effective one-dimensional potential. The simplest way to model the vacuum-metal interface is by assuming a \emph{discontinuous square potential barrier}, a potential energy structure with a negative step of height equal to the inner metal potential $V_0$ situated at the planar surface boundary $x=0$.
\begin{equation*}
 V(x) =
 \begin{cases}
 -V_0\,, & x\leq 0 \\
 \,\,\,\,\,0\,, & x\geq 0
 \end{cases}\,\,\,\,\,\,\,.
\end{equation*}
The one-electron Schr\"odinger equation can be trivially solved for such a potential energy. However, the analytic solution yields a spurious $100\%$ reflection at the limit of very low incident energy. This is a consequence of the unphysical discontinuity at the interface\,\cite{reflec01}. Such a model neglects the electromagnetic response of the metal to the incoming charge, which leads to a force acting on the electron in a manner that guarantees that its potential energy varies continuously. The next simplest way that remedies this problem is to assume a \emph{continuous image potential barrier}, a potential energy structure that follows the classical electrostatics image law.
\begin{equation*}
 V(x) =\begin{cases}
 -V_0\,, & x\leq x_0 \\
 -e^2/(4x)\,, & x\geq x_0
 \end{cases}\,\,\,\,\,\,\,,
\end{equation*}
where $x_0$ is the displacement of the image charge plane with respect to the surface of the metal, given by $x_0=e^2/(4V_0)$ in order to ensure continuity. The one-electron Schr\"odinger equation can be analytically solved for such a potential energy\,\cite{reflec02}. It yields a reflection coefficient that never exceeds $0.07$ at the limit of zero incident energy and behaves as a monotonically decreasing function of the incident energy. The results are similar\,\cite{reflec01} for more involved and more accurate image potentials of quantum mechanical origin\,\cite{reflec03,reflec04,reflec05}. However, such a model neglects the periodic nature of the potential in crystals. A more elaborate model that considers the metal interior is the \emph{continuous image-sinusoidal potential barrier} that superimposes a sinusoidal modulation for $x\leq{x}_0$.
\begin{equation*}
 V(x) =\begin{cases}
 -V_0+V_1\sin{\left[a(x-x_0)\right]}\,, & x\leq x_0 \\
 -e^2/(4x)\,, & x\geq x_0
 \end{cases}\,\,\,\,\,\,\,,
\end{equation*}
where $a$ and $V_1$ are adjustable parameters. The one-electron Schr\"odinger equation can also be analytically solved for such a potential energy\,\cite{reflec06}. It yields a reflection coefficient that contains incident energy bands where the reflection reaches $100\%$. This is a consequence of the so-far neglected inelastic component of the problem, since for instance the incoming electron can lose part of its incident total energy by interacting with the valence electrons of the metal\,\cite{reflec07}. Essentially, the problem is of many-body nature and cannot be described by a real effective potential. However, a \emph{complex potential barrier} can be assumed, where the imaginary part describes the inelastic aspects\,\cite{reflec08,reflec09}. The one-electron Schr\"odinger equation for such a potential no longer preserves the probability current, which simply implies scattering of the electron to another energy\,\cite{reflec10}. There only exist numerical solutions for such model potentials. In accordance with experiments, the reflection peaks are broadened and get suppressed far below unity\,\cite{reflec11,reflec12}. Moreover, reflection is no longer purely elastic but also bears an inelastic component, again as measured in experiments.

\begin{figure}
\centering
\includegraphics[width=3.3 in]{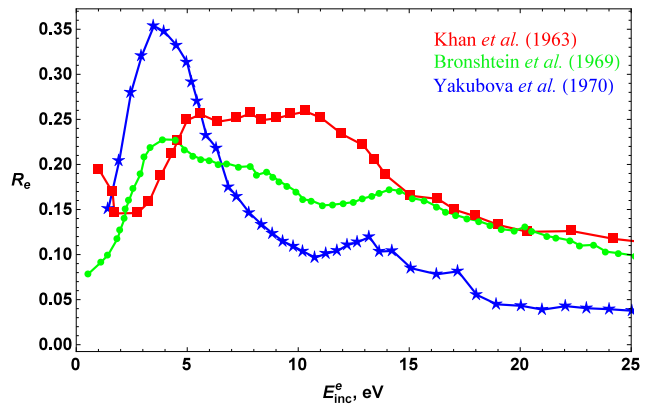}
\caption{The elastic electron reflection coefficient $R_{\mathrm{e}}(E_{\mathrm{inc}}^{\mathrm{e}})$ for normal electron incidence on monocrystalline tungsten - W(110) - with energies in the range $E_{\mathrm{inc}}^{\mathrm{e}}=0-25\,$eV. Results from the experiments of Bronshtein \emph{et al}\,\cite{experef1}, Khan \emph{et al}\,\cite{experef2} and Yakubova \emph{et al}\,\cite{experef3}.}\label{refl1}
\end{figure}

\begin{figure}
\centering
\includegraphics[width=3.3 in]{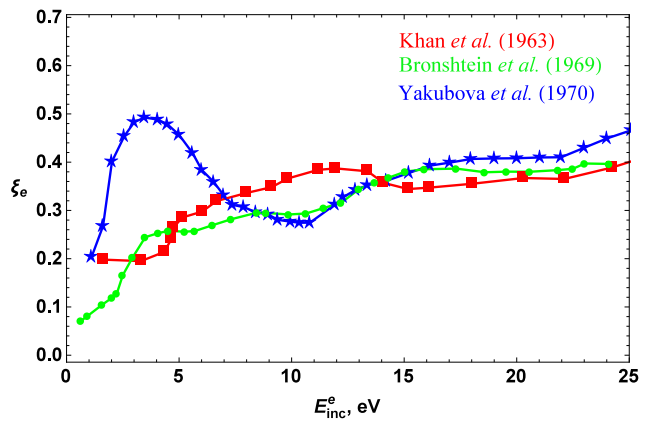}
\caption{The total electron reflection coefficient $\xi_{\mathrm{e}}(E_{\mathrm{inc}}^{\mathrm{e}})$ for normal electron incidence on monocrystalline tungsten - W(110) - with energies in the range $E_{\mathrm{inc}}^{\mathrm{e}}=0-25\,$eV. Results from the experiments of Bronshtein \emph{et al}\,\cite{experef1}, Khan \emph{et al}\,\cite{experef2} and Yakubova \emph{et al}\,\cite{experef3}.}\label{refl2}
\end{figure}

\noindent \emph{Survey of experimental results.} Clearly, the calculation of the reflection coefficient $\xi_{\mathrm{e}}$ from first principles is a formidable task. In addition, the functional form of $\xi_{\mathrm{e}}(E_{\mathrm{inc}}^{\mathrm{e}})$ is so highly non-monotonic and depends so strongly on the composition of the metal surface that simple empirical expressions are very hard to construct. Consequently, one has to resort to experimental results that are very sparse and contradicting\,\cite{experef1,experef2,experef3,experef4}. Fortunately, tungsten is the most studied material but almost exclusively as a single crystal (it is essential to have a well-defined surface for reproducibility). We point out that very low energy experiments are notoriously hard to perform. Apart from strict requirements in terms of surface conditions, detector design, beam collimation and stability, one needs to be confident that the collected electron current is indeed reflected from the metal target and is not part of the incident current that never reached the target\,\cite{critici2}. These difficulties are reflected in the strong deviations in the measured $\xi_{\mathrm{e}}(E_{\mathrm{inc}}^{\mathrm{e}})$ and $R_{\mathrm{e}}(E_{\mathrm{inc}}^{\mathrm{e}})$\,\cite{experef1,experef2,experef3,experef4} - in what follows we shall denote the total electron reflection coefficient by $\xi_{\mathrm{e}}$ and its elastic part by $R_{\mathrm{e}}$.

The experimental results of Bronshtein \emph{et al}\,\cite{experef1} and Khan \emph{et al}\,\cite{experef2} are considered as the most reliable in the literature. As seen in figure \ref{refl1}, in spite of the non-monotonic nature of the elastic electron reflection coefficient, they correlate very well with each other. As seen in figure \ref{refl2}, they exhibit less than $25\%$ deviations for the total electron reflection coefficient in the whole energy range with the exception of a very narrow region below $3\,$eV. On the contrary, the results of Yakubova \emph{et al}\,\cite{experef3} for $R_{\mathrm{e}}$ strongly deviate in the whole range, while for $\xi_{\mathrm{e}}$ they start deviating below $7\,$eV. Results from other authors exhibit much larger deviations\,\cite{experef4,experef5,experef6,experef7,experef8,experef9,experef0}, but the measured value of $\xi_{\mathrm{e}}$ for clean metal surfaces is always smaller than unity.

\begin{figure}
\centering
\includegraphics[width=3.3 in]{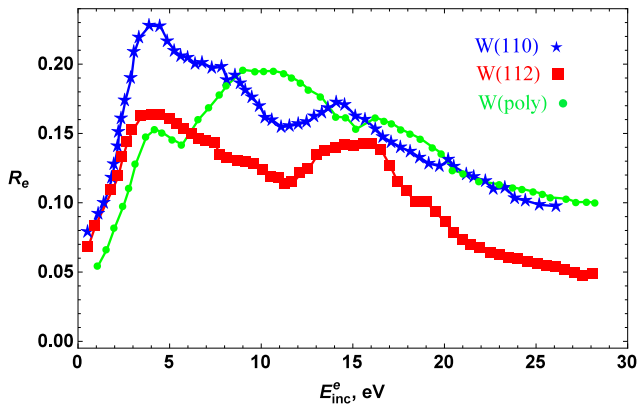}
\caption{The elastic electron reflection coefficient $R_{\mathrm{e}}(E_{\mathrm{inc}}^{\mathrm{e}})$ for normal incidence of electrons on tungsten with energies roughly in the range $E_{\mathrm{inc}}^{\mathrm{e}}=0-30\,$eV. Results from the experiments of Bronshtein \emph{et al}\,\cite{experef1} for polycrystalline tungsten and different tungsten single crystals, W(110) and W(112).}\label{refl3}
\end{figure}

\begin{figure}
\centering
\includegraphics[width=3.3 in]{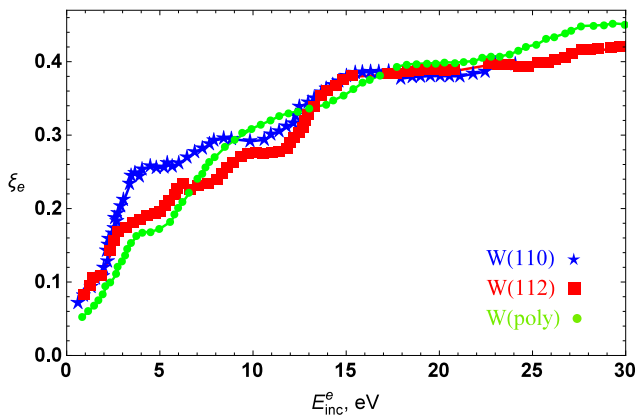}
\caption{The total electron reflection coefficient $\xi_{\mathrm{e}}(E_{\mathrm{inc}}^{\mathrm{e}})$ for normal incidence of electrons on tungsten with energies roughly in the range $E_{\mathrm{inc}}^{\mathrm{e}}=0-30\,$eV. Results from the experiments of Bronshtein \emph{et al}\,\cite{experef1} for polycrystalline tungsten and different tungsten single crystals, W(110) and W(112).}\label{refl4}
\end{figure}

The experiments of Khan \emph{et al}\,\cite{experef2} span the energy range of $0-100\,$eV for three different single crystal orientations and also study the effect of adsorbates. However, we recommend the use of the results of Bronshtein \emph{et al}\,\cite{experef1} for fusion applications, since these authors are the only ones that also employed polycrystalline tungsten. The experimental data are plotted in figures \ref{refl3} and \ref{refl4}. It is worth mentioning that the investigations of Bronshtein \emph{et al} are the most comprehensive\,\cite{experef1}; low energy electron reflection experiments were carried out with a large number of elemental metal surfaces (tungsten, nickel, aluminum, copper, molybdenum, silver, tantalum, gold), semi-conducting surfaces (silicon, germanium) but also with composite surfaces (oxides, salts). Independent of the nature of the surface studied, the measured value of $\xi_{\mathrm{e}}$ close to limit of zero incident energy was always smaller than unity.

With the aid of figure \ref{refl4}, we can demonstrate that the extrapolation of secondary electron emission yield formulas to very low energies apart from being physically wrong can also lead to misleading results. We shall employ the Young-Dekker formula for the tungsten values $E_{\mathrm{max}}=600\,$eV, $\delta_{\mathrm{\max}}=0.927$ and the optimal exponent $k=1.38$\,\cite{mypaper2}. For $E_{\mathrm{inc}}^{\mathrm{e}}=5\,$eV the extrapolated yield is $\delta\simeq0.02$ in contrast to $\xi_{\mathrm{e}}\simeq0.17$, for $E_{\mathrm{inc}}^{\mathrm{e}}=10\,$eV the extrapolated yield is $\delta\simeq0.04$ in contrast to $\xi_{\mathrm{e}}\simeq0.3$, for $E_{\mathrm{inc}}^{\mathrm{e}}=20\,$eV it is $\delta\simeq0.075$ compared to $\xi_{\mathrm{e}}\simeq0.4$. For very low energies, the difference between the extrapolated and the experimental values can even reach one order of magnitude!

\noindent \emph{Angle of incidence dependence.} To our knowledge, there are no experiments available that study the dependence of $\xi_{\mathrm{e}}$ and $R_{\mathrm{e}}$ on the angle of incidence. Based on the above theoretical considerations, one could assume that the problem is inherently one-dimensional. Therefore, only the normal velocity component would be relevant and a $E_{\mathrm{inc}}^{\mathrm{e}}\to{E}_{\mathrm{inc}}^{\mathrm{e}}\cos^2{\theta}$ mapping would be viable, where $\theta$ is the incident angle measured from the normal to the surface. This is not valid for polycrystalline tungsten due to electrostatic patch effects\,\cite{reflec07}. The general consensus in the literature is that the work function of polycrystalline tungsten at room temperature is $W_{\mathrm{f}}\simeq4.55\,$eV\,\cite{workfun1,workfun2,workfun3,workfun4}. However, the work function of monocrystalline tungsten exhibits strong variations for different crystallographic orientations\,\cite{workfun4}, for instance in the case of low index crystal faces $\mathrm{W}(110)\to5.30\,$eV, $\mathrm{W}(100)\to4.58\,$eV, $\mathrm{W}(111)\to4.40\,$eV. Therefore, the work function deviations between different crystal faces can reach $0.9\,$eV. This leads to a different surface potential on each mono-crystal region (polycrystalline tungsten surfaces are not strictly equipotential as stated by elementary electromagnetic theory) and consequently to strongly localized patch fields that can be important for surface effects such as low energy electron reflection. In absence of experimental results, we shall assume that there is a weak angle of incidence dependence based on the argument of Stangeby that the omnipresent surface roughness diminishes the importance of such an effect\,\cite{Stangeby}.

\noindent \emph{Magnetic field suppression of electron reflection.} Finally, we emphasize that the use of experimental results for the low energy electron reflection coefficient is only appropriate for devices with magnetic field orientation normal to the plasma-facing-component (PFC). For such a magnetic field topology, the gyromotion of the re-emerging electrons cannot lead to their recapture. However, in tokamaks, magnetic field lines connect to the PFCs nearly tangentially. In devices such as ITER, also due to the high magnetic field strength, the reflected electrons could promptly return to the PFC within their first gyration\,\cite{magsupp1,magsupp2,magsupp3}. The return fraction will be determined by the competition between the Lorentz force and the repelling electrostatic sheath force, thus it will strongly depend on the energy distribution of the reflected electrons. Undoubtedly, low energy reflection will be suppressed but the degree of suppression can only be quantified by simulations (particle-in-cell codes), where the data quoted above can serve as an input. Two additional comments are in order: First, since reflected electrons nearly retain the incident electron energy, they can be much more energetic than secondary electrons or thermionic electrons\,\cite{mypaper1}. Secondary electrons can be assumed to follow the Chung-Everhart energy distribution\,\cite{ChungEve}, which leads to a most probable energy ${W}_{\mathrm{f}}/3\simeq1.5\,$eV. Thermionic electrons can be assumed to follow a Maxwellian energy distribution with temperature equal to the surface temperature ${T}_{\mathrm{s}}$ of the PFC\,\cite{reflec07}, which leads to a most probable energy ${T}_{\mathrm{s}}/2<0.3\,$eV. This implies a stronger Lorentz force and a more effective magnetic suppression of reflected electrons\,\cite{magsupp4,magsupp5,magsupp6}. Second, it has never been acknowledged in the literature that the return fraction from any emission process will not be fully absorbed by the PFC but part of it will be reflected. This highlights the fact that without the aid of simulations it is neither possible to determine the steady state characteristics of this local electron population (energy distribution, number density) nor to quantify its effect on the global sheath structure.

\end{document}